\documentclass[aps,twocolumn,pra,superscriptaddress,showpacs,tightenlines]{revtex4}
\usepackage{tipa}
\usepackage{bbm}
\usepackage{amssymb}
\usepackage{amsmath}
\usepackage{graphicx}
\usepackage{epsfig}
\usepackage{txfonts}
\usepackage{subfigure}
\usepackage{amsfonts}
\usepackage{CJK}
\begin{document}

\begin{CJK*}{GBK}{song}

\title{Controlling decoherence speed limit of a single impurity atom \\in a Bose-Einstein-condensate reservoir}

\author{Ya-Ju Song}
\affiliation{\textit{Key Laboratory of Low-Dimensional Quantum
Structures and Quantum Control of Ministry of Education, Department
of Physics and Synergetic Innovation Center for Quantum Effects and
Applications, Hunan Normal University, Changsha 410081, China}} % Author affiliation

\affiliation{\textit{Department of Physics and Electronic
Information
Science, Hengyang Normal University, Hengyang 4210002, China}} % Author affiliation

\affiliation{\textit{Graduate School of Chinese
Academy of Engineering Physics, Beijing 10084, China}} % Author affiliation

\author{Le-Man Kuang\footnote{Author to whom any correspondence should be
addressed. }\footnote{ Email: lmkuang@hunnu.edu.cn}}
\affiliation{\textit{Key Laboratory of Low-Dimensional Quantum
Structures and Quantum Control of Ministry of Education, Department
of Physics and Synergetic Innovation Center for Quantum Effects and
Applications, Hunan Normal University, Changsha 410081, China}} % Author affiliation

\date{\today}

\begin{abstract}
We study  the decoherence speed limit (DSL) of a single impurity
atom immersed in a Bose-Einstein-condensed (BEC) reservoir when the
impurity atom is in a double-well potential.  We demonstrate how the
DSL of the impurity atom can be manipulated by engineering  the BEC
reservoir and  the impurity potential  within experimentally
realistic limits. We show that the DSL  can be controlled by
changing key parameters such as the condensate scattering length,
the effective dimension of the BEC reservoir, and the spatial
configuration of the double-well potential imposed on the impurity.
We uncover the physical mechanisms of controlling the DSL at root of
the spectral density of the  BEC reservoir.
\end{abstract}

%03.65.Ta: Foundations of quantum mechanics
%03.67.Lx   Quantum computation architectures and implementations
%03.67.-a: Quantum information
%03.65.Yz: Decoherence; open systems; quantum statistical methods
 \maketitle \narrowtext
\end{CJK*}
\section{\label{Sec:1}Introduction}
Quantum coherence is the essential reason for quantum
counterintuitive features that challenge our classical perception of
nature, and long coherence time is a crucial condition for the
viability of performing quantum information processing. However, the
unavoidable couplings between quantum systems and their reservoirs
induce the phenomenon of quantum decoherence \cite{Breuer2002}.
Developing a quantitative understanding of the decoherence mechanism
and exploring controllable methods of decoherence \cite{Leggett1987,Prokof2000,Kuang1999-1,Kuang1995,Kuang1999-2,Liao2011,Yuan2010,Lu2010,He2017} are therefore critical.
In recent years, much attention has been paid to the realization
of long-lived quantum coherence. But few works focus on the lower
bound of coherence time, and how to regulate the lower bound of
coherence time still remains an open question. The aim of this paper
is to address this problem making use of quantum speed limit (QSL) for a single impurity atom in a Bose-Einstein-condensed (BEC) reservoir.

The QSL time, denoted by $\tau_{QSL}$, is defined as the minimal
time between two distinguishable states of a quantum system. It can
be used to characterize the ultimate bound imposed by quantum
mechanics on the maximal evolution speed. Recently, based on various
distance metrics of two distinguishable states or the notion of
quantumness, different bounds on the $\tau_{QSL}$ for both isolated
\cite{Mandelstam1945,Margolus1998,Vittorio2003} and open
\cite{Taddei2013,Campo2013,Sebastian2013,Deffner02017,Diego2016,Cimmarusti2015,Jones2010,Zhang2014,Sun2015,Mirkin2016,Xu2016,Meng2015,Song2016,Song2017,Deffner2017,Campaioli2018,Mo2017,Lee2018}
system dynamics have been obtained. In this paper, we adopt the
$\tau_{QSL}$ derived by Taddei $\emph{et al}$, who use quantum
Fisher information for time estimation and choose Bures fidelity as
the distance measure between two quantum states. For a given
distance, a shorter $\tau_{QSL}$ implies a higher dynamical speed
upper bound, otherwise, longer $\tau_{QSL}$ means a lower dynamical
speed limit. As for a quantum pure dephasing model, the $\tau_{QSL}$
is exactly the lower bound of coherence time, and it can be used to
characterize the upper bound of quantum decohering speed, i.e., the
quantum decoherence speed limit (DSL).

Here we consider a single impurity atom immersed in a BEC. The
impurity atom is confined by a deep, symmetric double well
potential, while the BEC atoms are confined by a very shallow
trapping potential. In
Refs.~\cite{Bloch2008,Cirone2009,Haikka2011,Klein2007,Yuan2017}, it
has been proved that the BEC atoms can be used to simulate a phase
damping reservoir for the doped impurity atom. Meanwhile, BEC
systems are essentially macroscopic quantum many body systems with
effectively controllable  dimension and nonlinear interaction. These
provide us with a platform to control the decoherence of the single
impurity atom by the use of the nonlinear BEC reservoir with
different dimensions. Here we mainly consider the effect of these
controllable parameters of the BEC reservoir on the impurity's lower
bound of coherence time, i.e., quantum DSL. We show that not only
the nonlinearity and the dimension of the BEC, but also the spatial
form of the double well potential imposed on the impurity can be all
treated as controllable parameters to manipulate the impurity's
lower bound of coherence time. And in order to insight the physical
mechanism behind the control, we also analyze the spectral density
in details. We find that the nonlinear interaction and the effective
dimension of the BEC can change the spectral density from a soft
subohmic spectrum to a hard superohmic spectrum, while the
characteristic length of each well and the distance between two
wells of the double well potential change the cut-off frequency and
the effective coupling constant in the spectral density,
respectively.

The paper is organized as follows. In Sec.~II, we introduce our
physical model and obtain the dynamics of the impurity in the
dephasing BEC reservoir. In Sec.~III, we investigate the possibility of manipulating the  quantum DSL of the impurity atom by  various controllable parameters such as the nonlinearity and the effective dimension of
the BEC reservoir, and the spatial parameters of double well
potential imposed on the impurity.  Finally, Sec.~IV is devoted to some
conclusions.

%%%%%%%%%%%%%%%%%%%%%%%%%%%%%%%%%%%%%%%%%%%%%%%%%%%%%%%%%%%%%%%%
\section{\label{Sec:2} Physical model and decoherence analysis}
%%%%%%%%%%%%%%%%%%%%%%%%%%%%%%%%%%%%%%%%%%%%%%%%%%%%%%%%%%%%%%%%

%%%%%%%%%%%%%%%%%%%%%%%%%%%%%%%%%%%%%%%%%%%%%%%%%%%%%%
\begin{figure}[tbp]
\includegraphics[clip=true,width=0.45\textwidth]{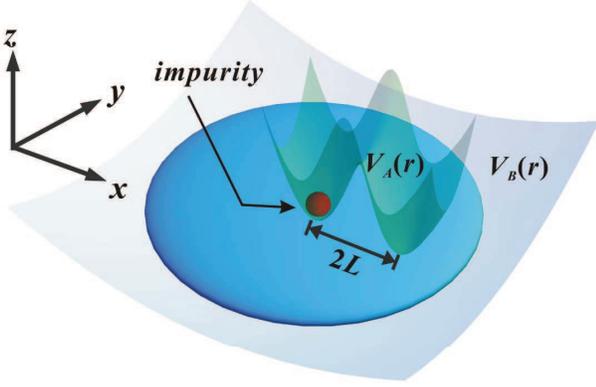}
\caption{Schematic of the impurity coupled to a
quasi-two-dimensional BEC environment. The impurity atom (red
circle) is trapped in a double well potential $V_{A}(\textbf{r})$
with the distance between two wells $2L$, while the BEC (blue
region) is confined in a shallow harmonic trap $V_{B}(\textbf{r})$.}
\label{1.eps}
\end{figure}
%%%%%%%%%%%%%%%%%%%%%%%%%%%%%%%%%%%%%%%%%%%%%%%%%%%%%%%%%%%%%%%%%%%%

We consider an impurity atom immersed in a BEC. As shown in Fig.~\ref{1.eps}, the  BEC  is confined by a very shallow trapping
potential $V_{B}(\textbf{r})$, while the impurity atom is trapped by
a deep, symmetric double well potential $V_{A}(\textbf{r})$ with the
distance between two wells $2L$. Here the occupations of the
impurity in the left and the right well represent two pseudo-spin
states, denoted by $|g\rangle\equiv|L\rangle$ and
$|e\rangle\equiv|R\rangle$, respectively. Assume that the double well is separated by
a high-energy barrier, the tunneling between the two wells can be
neglected. At low energies, only the contact interaction between the
impurity and the BEC atoms contributes significantly. Then the BEC
atoms could serve as a ground for simulating a phasing damping
environment for the doped impurity atom. The Hamitonian of the impurity-BEC system
takes the form of an effective spin-boson model
\cite{Cirone2009,Haikka2011} ($\hbar=1$),
\begin{eqnarray}
\label{eq-1}
\hat{H}_{\rm{eff}}&=&\omega _{0}\hat{\sigma}_{z}+\sum_{\mathbf{k}}\omega _{\mathbf{k}}\hat{a}_{\mathbf{k}%
}^{\dagger }\hat{a}_{\mathbf{k}}+\hat{\sigma}%
_{z}\sum_{\mathbf{k}}\left( \xi_{\mathbf{k}}\hat{a}_{\mathbf{k}}^{\dagger }+\xi_{%
\mathbf{k}}^{\ast }\hat{a}_{\mathbf{k}}\right),
\end{eqnarray}
where $\sigma_{z}=|R\rangle\langle R|-|L\rangle\langle L|$, $\omega
_{0}$ is the effective energy difference between the two pseudo-spin
states of the impurity, and $\hat{a}_{\mathbf{k}}^{\dagger}
(\hat{a}_{\mathbf{k}})$ is the creation (annihilation) operator of
the Bogoliubov phonon of the BEC on the top of condensate wave
function with the energy $\omega _{\mathbf{k}}$
\cite{Lifshitz1980,Ohberg1997}, and $\xi_{\mathbf{k}}$ is the
coupling constant between the impurity and the BEC.

The BEC excitations obey the following dispersive relation

\begin{eqnarray}
\label{eq-2}
\omega _{\mathbf{k}} &=&\sqrt{2\epsilon _{\mathbf{k}%
}n_{D}g_{D}+\epsilon _{\mathbf{k}}^{2}},
\end{eqnarray}
where the subscript $D$ denotes the effective dimension of the BEC,
$n_{D}$ is the condensate number density, $g_{D}$ is the
inter-atomic nonlinear coupling constant, and $\epsilon
_{\mathbf{k}}=\frac{k^{2}}{2m_{B}}$ is the free-particle energy with
the mass of a background gas particle $m_{B}$ and the momentum
$k=|\mathbf{k}|$. It is worth noting that from Eq. (2) we can obtain
two interesting limit cases of the dispersive relation
\cite{Cirone2009,Zhou,Guo}. The first is the phonon-like type
dispersive relation $\omega _{\mathbf{k}} \propto k$,  which happens
in the regime of low-energy excitations $ \epsilon
_{\mathbf{k}}\ll2n_{D}g_{D}$  or  strong nonlinear interaction
$g_{D}\gg\epsilon _{\mathbf{k}}/2n_{D}$. The second is the
free-particle type dispersive relation $\omega _{\mathbf{k}} \propto
\epsilon _{\mathbf{k}}$, which occurs in the regime of high-energy
excitations or when the inter-atomic nonlinear interaction can be
ignored.

Under the harmonic approximation through calculating the
analytical expression of the ground state wave function of the
impurity in each well, we can get the coupling constant between the impurity and the BEC \cite{Cirone2009,Haikka2011}
\begin{eqnarray}
\label{eq-3}
\xi_{\mathbf{k}}=-i\eta _{D}\sqrt{\frac{n_{D}\epsilon _{\mathbf{k}}}{V\omega _{%
\mathbf{k}}}}e^{-k^{2}\sigma ^{2}/4}\sin \left( \mathbf{k}\cdot \mathbf{L}%
\right) e^{-i\mathbf{k}\cdot \mathbf{r}_{W}},
\end{eqnarray}
where $\eta_{D}$ is the coupling constant of the impurity-BEC
contact interaction with volume $V$, $\mathbf{r}_{W}$ is the center
coordinate of the double well potential $V_{A}(\textbf{r})$, and
$\sigma=\frac{\hbar}{\sqrt{m_{A}\omega}}$ is the characteristic
length of two approximate harmonic wells of $V_{A}(\textbf{r})$ with
the trapping frequency $\omega$ and the mass of the impurity
$m_{A}$.

Note that, by changing the shape of the trapping potentials, it is
possible to produce dilute gases in highly anisotropic
configurations, where the motion of BEC atoms is quenched in
quasi-one-dimensional (1D) or quasi-two-dimensional (2D) directions
\cite{Pitaevskii2003,Naidon2007,Hangleiter2015,Naidon2006,Catani2012,Bloch2008}.
The consequent inter-atomic interaction strengths ( $g_{1}$ and
$g_{2}$) in 1D and 2D BEC can be expressed in terms of the
inter-atomic interaction  strength ($g_{3}$) in 3D BEC
\begin{equation}
g_{1}=\frac{g_{3}}{2\pi a_{\perp ,B}^{2}}, \hspace{0.5cm} g_{2}=\frac{g_{3}}{\sqrt{2\pi }a_{z,B}},   \hspace{0.5cm}
g_{3}=\frac{4\pi a_{B}}{m_{B}},
\end{equation}
where $a_{B}$ is the  tunable $s$-wave scattering length of the BEC.
$a_{\perp,B }$ and $a_{z,B}$ are  the transversal width and the
axial length of the wave function of the BEC atoms, respectively.

Similarly, one can obtain the number densities of the 1D and 2D BEC
with $n_{3}$ being the number density of the 3D BEC
\cite{Pitaevskii2003,Bloch2008}
\begin{equation}
n_{1}=\pi n_{3} a_{\perp ,B}^{2}, \hspace{0.5cm}
n_{2}=\sqrt{\pi}n_{3}a_{z,B}.
\end{equation}

And the 1D and 2D impurity-BEC coupling constants are expressed as
\begin{equation}
\eta_{1}=\frac{\eta_{3}}{\pi( a_{\perp,A}^{2}+a_{\perp,B}^{2})},
\eta_{2}=\frac{\eta_{3}}{\sqrt{\pi(a_{z,A}^{2}+a_{z,B}^{2})}},   \eta_{3}=\frac{2\pi a_{AB}}{m_{AB}},
\end{equation}
where $a_{\perp,A}$ is the transversal width of $V_{A}(\textbf{r})$
for the 1D case, and $a_{z,A}$ is the axial length for the 2D case.
$a_{AB}$ is the $s$-wave scattering length for impurity-BEC
collisions and $m_{AB}=\frac{m_{A} m_{B}}{m_{A}+m_{B}}$ is  the
reduced mass.

For the sake of simplicity, we consider the case that the BEC reservoir is at
zero-temperature. Assume that the qubit is initially in an
arbitrary state $\hat{\rho}\left( 0\right)= \frac{1}{2}(\mathbb{I}+
x\sigma _{x}+ y\sigma _{y}+z \sigma _{z})$. In the interaction
picture with respect to $\hat{H}_{0}=\omega
_{0}\hat{\sigma}_{z}+\sum_{\mathbf{k}}\omega
_{\mathbf{k}}\hat{a}_{\mathbf{k} }^{\dagger }\hat{a}_{\mathbf{k}}$,
the exact reduced impurity dynamics can be obtained by using Magnus
expansion \cite{Blanes2009} with the following expression
\begin{equation}
\label{eq-4} \hat{\rho}\left( t\right) =\frac{1}{2}\left(
\begin{array}{cc}
1+z & (x-iy)e^{-\Gamma\left( t\right) } \\
(x+iy)e^{-\Gamma\left( t\right) } & 1-z%
\end{array}
\right),
\end{equation}
where the dephasing function $\Gamma\left( t\right) $
\cite{Cirone2009,Haikka2011} is given by,
\begin{eqnarray}
\label{eq-5} \Gamma \left( t\right) &=& \frac{8\eta
_{D}^{2}n_{D}}{\left( 2\pi \right) ^{D}}\int dk k^{D-1} f_{D}(k L)
\frac{e^{-k^{2}\sigma ^{2}/2}\sin ^{2}\left( \omega
_{\mathbf{k}}t/2\right) }{\omega _{\mathbf{k}}\left(
2n_{D}g_{D}+\epsilon _{\mathbf{k}}\right) }.
\end{eqnarray}
During the calculation of $\Gamma(t)$, we have used the continuum
limit $\frac{1}{V}\sum_{\mathbf{k}} \rightarrow \frac{1}{\left( 2\pi
\right) ^{D}}\int d\Omega _{D}\int dk k^{D-1}$. The angular integral
is defined as $f_{D}(k L)=\int d\Omega _{D}\sin ^{2}\left(
\mathbf{k}\cdot \mathbf{L}\right)$ with $\Omega _{D}$ being the
surface of the unit sphere in $D$ dimensions. It is not difficult to
find that the angular integrals read as $f_{1}(k L)= \sin ^{2}(kL)
$, $f_{2}(k L)=\pi[1-J_{0}(2kL)]$ with $J_{0}(x)$ being the Bessel
function of the first kind, and $f_{3}(k
L)=2\pi[1-\frac{\sin(2kL)}{2kL}]$. Up to now, the dephasing function
of the impurity atom in the $D$-dimensional BEC environment can be
obtained by combining Eqs.~(\ref{eq-2}),(\ref{eq-5}). Significantly,
the dephasing function $\Gamma(t)$ is not only related to the
impurity-BEC coupling constant $\eta_{D}$, but also can be modulated
by the boson-boson s-wave scattering length $a_{B}$ through $g_{D}$,
the effective dimension $D$ of the BEC reservoir, as well as the
spatial form of the double well potential $V_{A}(\textbf{r})$,
including the parameters $\sigma$ and $L$.

%%%%%%%%%%%%%%%%%%%%%%%%%%%%%%%%%%%%%%%%%%%%%%%%%%%%%%%%%%%%%%%%%%%%%%
\begin{figure}[tbp]
\centering
\includegraphics[clip=true,width=0.4\textwidth]{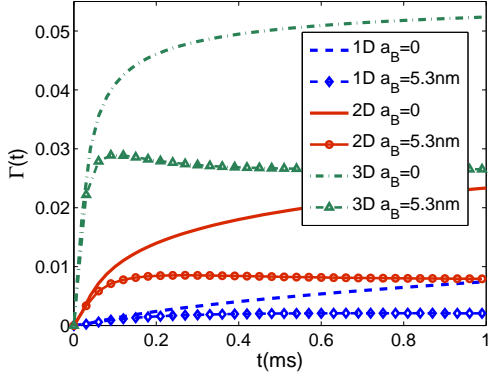}
\caption{(Color online) Comparison of the dynamical behaviors of the
dephasing function $\Gamma(t)$ in free and interacting BEC
reservoirs with different dimensions. The blue dashed (with empty
rhombus) line represents the free (interacting) $1$D case, the red
solid (with empty circles) line represents the free (interacting)
$2$D case, and the green dot-dashed (with empty triangle) line
represents the free (interacting) $3$D case. Here the value of
$a_{B}$ takes its nature value without using Feshbach resonances,
$a_{B}=a_{NV}=5.3nm$, for the interacting case. Other parameters are
given by $n_{3}=10^{20}m^{-3}$, $m_{B}=14.45\times10^{-26}kg$,
$m_{AB}=3.02\times10^{-26}kg$, $a_{\bot,A}=a_{\bot,B}=100a_{NV}$,
$a_{z,A}=a_{z,B}=100a_{Rb}$, $\sigma=45nm$, $L=150nm$,
$a_{AB}=55a_{0}$ with the Bohr radius $a_{0}$.}\label{2.eps}
\end{figure}
%%%%%%%%%%%%%%%%%%%%%%%%%%%%%%%%%%%%%%%%%%%%%%%%%%%%%%%%%%%%%%%%%%%%%%

Fig.~\ref{2.eps} indicates the dependence of the dephasing function
$\Gamma(t)$ on the effective dimension $D$ and the scattering length
$a_{B}$ of the BEC atoms. Here the kind of atoms for the impurity is
$^{23}Na$ with the mass $m_{A}\approx3.82\times10^{-26}kg$, while
the kind of atoms for the BEC is $^{87}Rb$ with the mass
$m_{B}\approx14.45\times10^{-26}kg$. One can create a
low-dimensional background BEC by a suitable modification of the
potential $V_{B}(\textbf{r})$. And the scattering length $a_{B}$ of
the background $^{87}Rb$ condensate gas can be tuned via Feshbach
resonances. Other parameters are given in the legend of
Fig.~\ref{2.eps} \cite{Cirone2009,Haikka2011,Jaksch1998}. As shown
in the Fig.~\ref{2.eps}, for the case of free bosons in BEC, i.e.,
$a_{B}=0$, the $\Gamma(t)$ monotonically increases with the time in
both the 1D and 2D BEC reservoirs. However, the $\Gamma(t)$ first
increases and then trends to a constant value in the 3D BEC
reservoir. These results indicate that the impurity would completely
dephase in the long-time limit for both the 1D and 2D cases, but
there exists a stationary coherence for the 3D case. As for the case
of presence of inter-atomic interaction in the BEC reservoir, where
the scattering length $a_{B}$ takes its natural value
$a_{Rb}\approx5.3nm$, the $\Gamma(t)$ falls down after it ascends
first, and finally tends to be stable for three kinds of dimensions.
That is to say, the inter-boson interaction or the nonlinearity of
BEC reservoir induces the appearance of the stationary coherence of
the impurity. And by comparing the dynamics of $\Gamma(t)$ in
different dimension cases, we can find that the coherence of the
impurity would also be enhanced by decreasing the effective
dimension of the BEC reservoir.

%%%%%%%%%%%%%%%%%%%%%%%%%%%%%%%%%%%%%%%%%%%%%%%%%%%%%%%%%%%%%%%%%%%%%%
\begin{figure}[tbp]
\centering
\includegraphics[clip=true,width=0.4\textwidth]{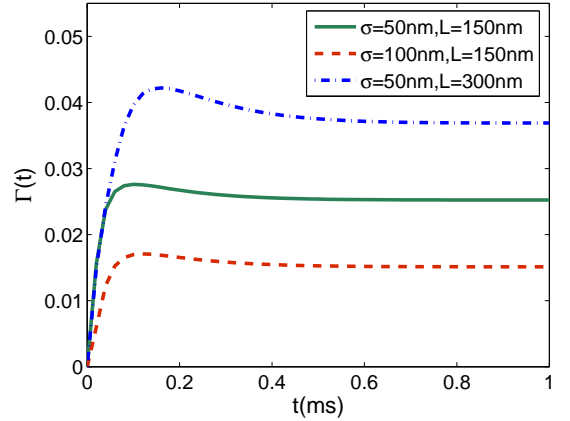}
\caption{(Color online) The dynamics of dephasing function
$\Gamma_{3}(t)$ in $3$D BEC reservoir for different spatial
parameters of the double-well. The green solid line represents
$\sigma=50nm, L=150nm$, the red dashed line represents
$\sigma=100nm, L=150nm$, and the blue dot-dashed line represents
$\sigma=50nm, L=300nm$. The scattering length of the background BEC
takes its nature value of $a_{B}=a_{NV}=5.3nm$. Other parameters are
the same as in Fig.~\ref{2.eps}.}\label{3.eps}
\end{figure}
%%%%%%%%%%%%%%%%%%%%%%%%%%%%%%%%%%%%%%%%%%%%%%%%%%%%%%%%%%%%%%%%%%%%%%

We now turn to investigate the influence of the spatial configuration of the
double well $V_{A}(\textbf{r})$, including the parameters $\sigma$
and $L$, on the dephasing function $\Gamma(t)$. Here
$\sigma=\sqrt{\frac{\hbar}{m_{A}\omega}}$ is the characteristic
length with $\omega$ being the trapping frequency of the harmonic
trap approximating the lattice potential at bottom of the left and
the right wells. $L$ is the half of the distance between two wells
of $V_{A}(\textbf{r})$. Both of them can be adjusted experimentally.
Inspecting Fig.~\ref{3.eps} we can find that for a given
inter-atomic interaction $a_{B}$, the dephasing function in the $3$D
BEC reservoir $\Gamma_{3}(t)$ decreases with increasing $\sigma$ or
shortening $L$. This means that the larger $\sigma$ or smaller $L$
could induce the slower dephasing speed, and we find these results
are also applicative for the low-dimension cases. In addition,
comparing Figs.~\ref{2.eps} with ~\ref{3.eps} one can conclude that
the stationary coherence is much sensitive to the nonlinearity
parameter $a_{B}$ of the BEC reservoir rather than the spatial
parameters $\sigma$ and $L$ of the double well $V_{A}(\textbf{r})$.
Anyway, these results provide us a way to control the quantum
dephasing speed of the impurity atom in its BEC reservoir.

%%%%%%%%%%%%%%%%%%%%%%%%%%%%%%%%%%%%%%%%%%%%%%%%%%%%%%%%%%%%%%%%
\section{\label{Sec:3} control of the decoherence speed limit}
%%%%%%%%%%%%%%%%%%%%%%%%%%%%%%%%%%%%%%%%%%%%%%%%%%%%%%%%%%%%%%%%

In this section, we show how to manipulate the DSL of the impurity atom through changing controllable parameters of the impurity atom and the BEC reservoir. As was mentioned in the
introduction section, the DSL can be characterized by the QSL time $\tau _{QSL}$. For a
given distance $\mathcal{D}$ between two distinguishable states, a longer $\tau
_{QSL}$ indicates a slower decohering limit speed, which means the
greater robustness of the impurity against dephasing induced by the
interaction with the BEC reservoir.
%And $\tau_{QSL}=\tau $ implies the qubit coherence decays on a geodesic.
%%%%%%%%%%%%%%%%%%%%%%%%%%%%%%%%%%%%%%%%%%%%%%%%%%%%%%%%%%%%%%%%%%%%%%

We now consider the DSL of the impurity atom between two quantum states with a distance $\mathcal{D}$.
The authors in Ref.~\cite{Taddei2013} have proved that the distance
between the initial states $\hat{\rho}(0)$ and the final states
$\hat{\rho}(\tau )$ is bounded by
\begin{equation}
\label{3eq-1}
\mathcal{D}\equiv \arccos \sqrt{F_{B}\left[ \hat{\rho}(0),\hat{\rho}(\tau )%
\right] }\leq \frac{1}{2}\int_{0}^{\tau }\sqrt{\mathcal{F}_{Q}(t)}\
dt,
\end{equation}
where the distance between two states is defined based on the Bures
fidelity $F_{B}\left[ \hat{\rho}(0),\hat{\rho}(\tau )\right] =\left[
\mathrm{tr}\sqrt{\hat{\rho}(0)^{1/2}\hat{\rho}(\tau)\hat{\rho}(0)^{1/2}}\right]
^{2}$, and $\mathcal{F}_{Q}(t)$ is the quantum Fisher information with respect to the time which
can be expressed as \cite{Helstrom1976}
\begin{equation}
\label{3eq-2} \mathcal{F}_{Q}\left( t\right)
=\sum_{m,n=\pm}\frac{2}{p_{m}+p_{n}} \left\vert \left\langle \psi
_{m}\right\vert \frac{\partial \hat{\rho} (t)} {\partial
t}\left\vert \psi _{n}\right\rangle \right\vert^{2}.
\end{equation}
where $p_{\pm}$ ($\left\vert \psi _{\pm}\right\rangle$) are the
eigenvalues (eigenstates) of the quantum state of the impurity atom.
For the density operator of the impurity atom given by
Eq.~(\ref{eq-4}) we have
\begin{eqnarray}
\label{3eq-3}
\left\vert \psi _{\pm}\right\rangle &=&N_{\pm} \left[(x-iy)e^{-\Gamma(t)}\left\vert e\right\rangle \mp (A\pm z)\left\vert g\right\rangle\right], \nonumber \\
 p_{\pm}&=&\frac{1}{2}\left( 1\mp A\right)
\end{eqnarray}
where we have introduced the following functions
\begin{eqnarray}
\label{3eq-5} N_{\pm}^{-2}=2A(A \pm z),  \hspace{0.5cm}
A=\sqrt{(x^{2}+y^{2})e^{-2\Gamma\left(t\right)}+z^{2}}.
\end{eqnarray}

Submitting Eqs.~(\ref{3eq-3}) and (\ref{3eq-5}) into Eq.~(\ref{3eq-2}),
we can obtain the quantum Fisher information with respect to the
time,
\begin{eqnarray}
\label{3eq-6} \mathcal{F}_{Q}\left( t\right) =\frac{\left(
1-z^{2}\right)(x^2+y^2) \dot{
\Gamma}^{2}(t) e^{-2\Gamma \left( t\right) }}{%
1-z^{2}-(x^2+y^2)e^{-2\Gamma (t) }}.
\end{eqnarray}

Thus, the upper bound of the distance $\mathcal{D}_{UB}$ between two
distinguishable states can be obtained by inserting
Eq.~(\ref{3eq-6}) into Eq.~(\ref{3eq-1}),
\begin{eqnarray}
\label{3eq-7}
\mathcal{D}_{UB}=\frac{1}{2}\int_{0}^{\tau}\frac{ \sqrt{(1-z^{2})(x^{2}+y^{2})}
\left\vert \dot{\Gamma}(t) \right\vert e^{-\Gamma (t) }}{\sqrt{
1-z^{2}-(x^2+y^2)e^{-2\Gamma(t) }}}dt
\end{eqnarray}

From Eq.~(\ref{3eq-7}) we can see that the initial zero coherence
$C\left[\rho(0)\right] \equiv x^2+y^2=0$ would lead to
$\mathcal{D}_{up}=0$ at any time. In other words, the above bound
consistently guarantees that the eigenstates of $\sigma_{z}$ do not
evolve. Furthermore, assuming the initial states of the impurity
$\hat{\rho}(0)$ are pure states, the bound saturates, i.e.,
$\mathcal{D}=\mathcal{D}_{UB}$, if and only if the $\hat{\rho}(0)$
is on the equator of the bloch sphere with $x^2+y^2=1$ and the first
derivative of the dephasing function versus time $\dot{\Gamma}\left(
t\right)>0$ within the driving time $t\in[0,\tau]$, which can be
proved to be a Markovian process \cite{Taddei2013}. Noting that the
bound saturation implies the dephasing channel connecting two states along a
geodesic path. More importantly, Eq.~(\ref{3eq-7}) clearly shows
that the $\mathcal{D}_{UB}$ for a given driving time $\tau$ is not
only determined by the initial-state parameters ($x$, $y$, $z$), but
also determined by the dephasing function $\Gamma(t)$ in
Eq.~(\ref{eq-5}) and its first derivative versus time
$\dot\Gamma(t)$. In fact, the dependence of the $\mathcal{D}_{UB}$
on the effective dimension $D$ of the BEC, the boson-boson
scattering length $a_{B}$, and the spatial form of the double-well
potential (i.e., $\sigma$, $L$) is exactly from the dephasing
function $\Gamma(t)$.

%%%%%%%%%%%%%%%%%%%%%%%%%%%%%%%%%%%%%%%%%%%%%%%%%
\begin{figure}[tbp]
\centering
\subfigure{\includegraphics[clip=true,width=0.4\textwidth]{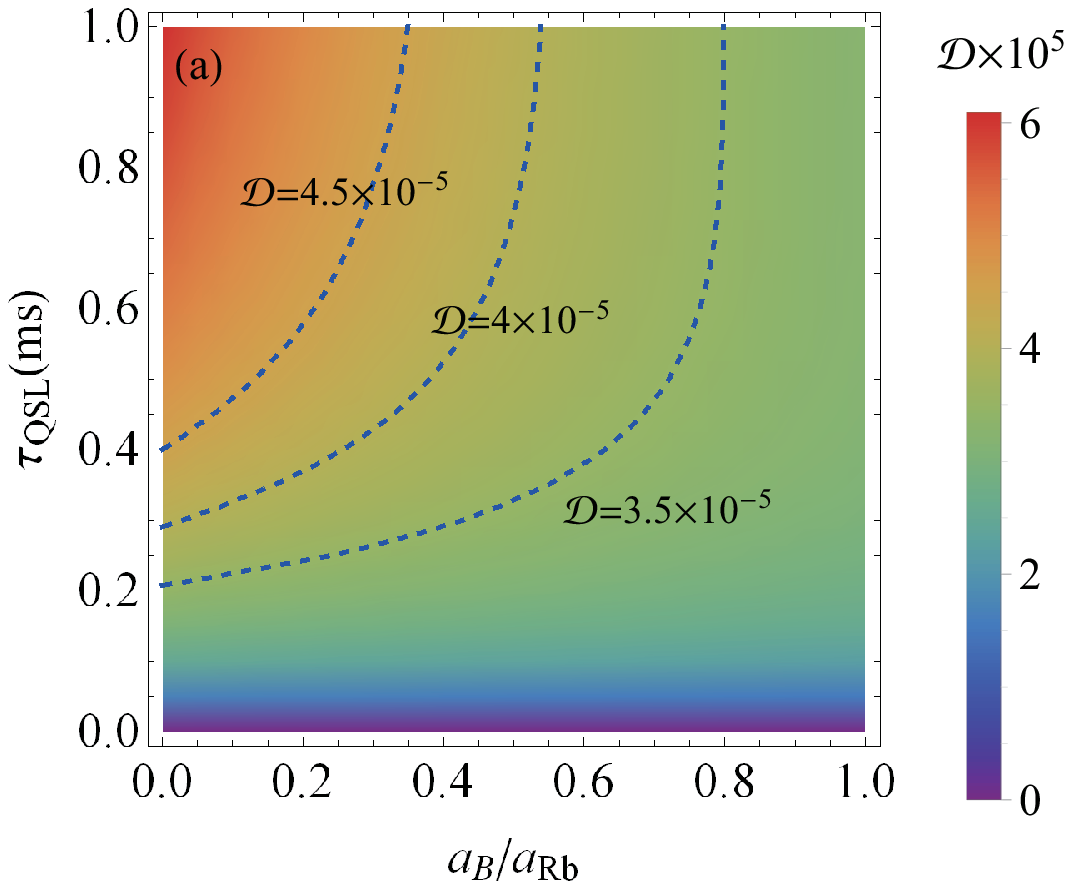}}
\subfigure{\includegraphics[clip=true,width=0.4\textwidth]{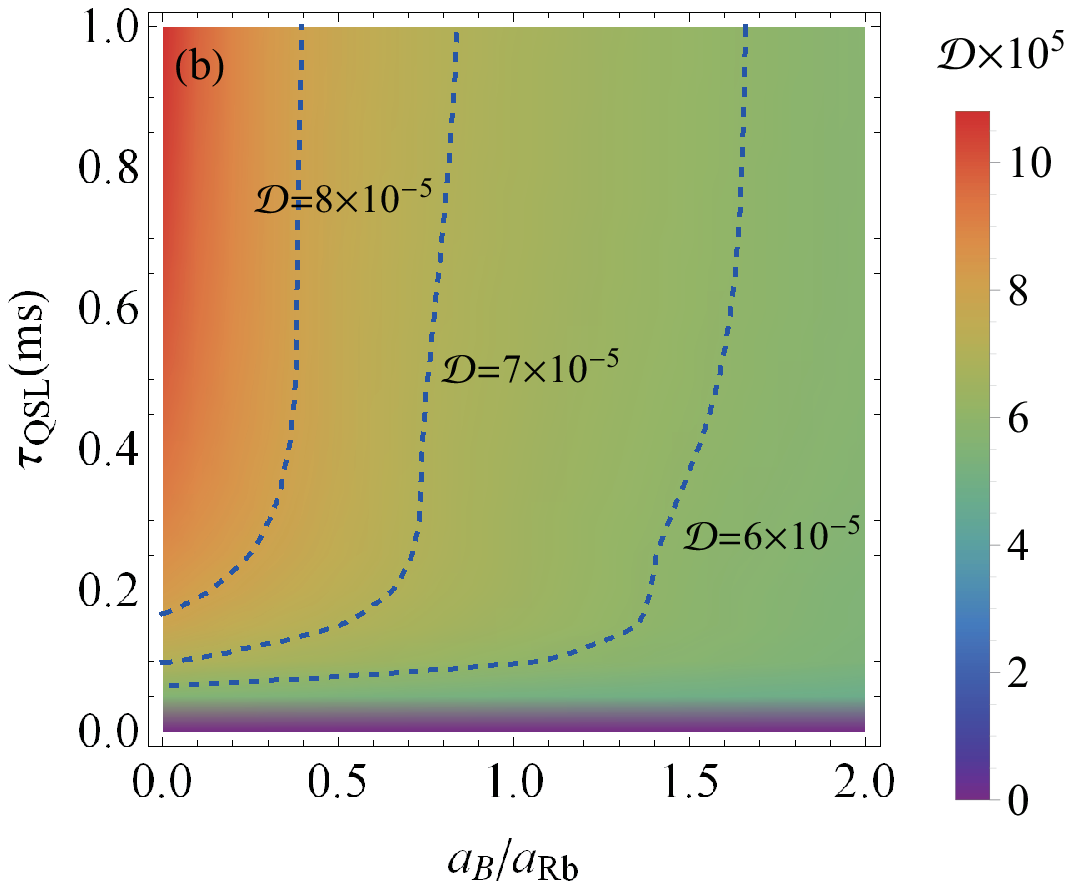}}
\subfigure{\includegraphics[clip=true,width=0.4\textwidth]{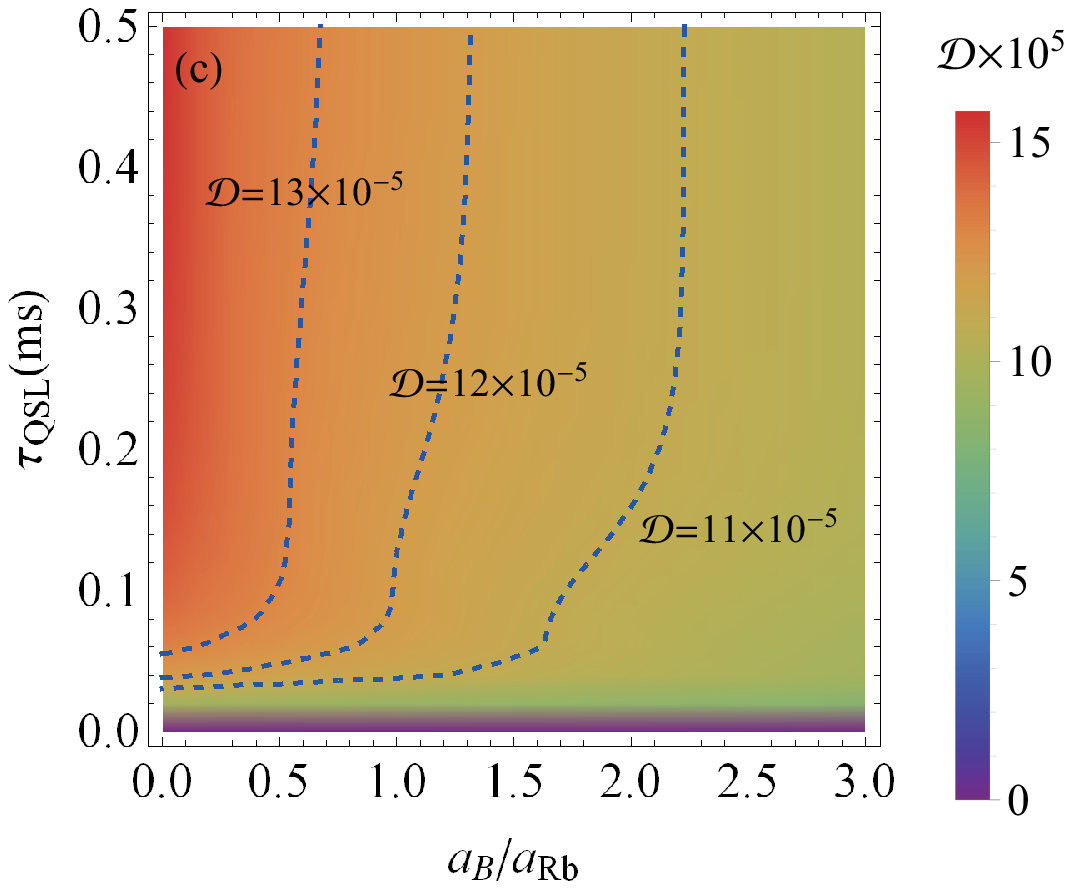}}
\caption{(Color online) The distance $\mathcal{D}$ as a function of
the scattering length $a_{B}$ and the QSL time $\tau_{QSL}$. The
blue dashed contour mark the dependence of the QSL time $\tau_{QSL}$
on the scattering length $a_{B}$ for given distances $\mathcal{D}$.
From top to bottom, the three subfigures represent (a) 1D case with
$\mathcal{D}\times10^{5}=4.5,4,3.5$, (b) 2D case with
$\mathcal{D}\times10^{5}=8,7,6$) and (c) 3D case with
$\mathcal{D}\times10^{5}=13,12,11$. The initial state of the
impurity is a maximally coherent states with $x^2+y^2=1$, $z=0$.
Other parameters are the same as in Fig.~\ref{2.eps}. }\label{4.eps}
\end{figure}
%%%%%%%%%%%%%%%%%%%%%%%%%%%%%%%%%%%%%%%%%%%%%%%%%%

The DSL is characterized by the QSL time $\tau _{QSL}$  between an
arbitrary initial state and a target state with a fixed distance
$\mathcal{D}$. It is the lower bound of the evolution time and  can
be derived through the following relation,
\begin{eqnarray}
\label{3eq-8} \mathcal{D}=\frac{1}{2}\int_{0}^{\tau
_{QSL}}\frac{\sqrt{(1-z^{2})(x^{2}+y^{2})} \left\vert
\dot{\Gamma}(t) \right\vert e^{-\Gamma (t) }}{\sqrt{
1-z^{2}-(x^2+y^2)e^{-2\Gamma(t) }}}dt.
\end{eqnarray}
Here $\tau _{QSL}$ servers a lower bound for the coherence time, and
it allows us to define a quantum dephasing speed limit (in frequency
units), $V_{QSL} = \frac{\mathcal{D}}{\tau _{QSL}}$
\cite{Mirkin2016}. For a given distance $\mathcal{D}$, a longer
$\tau _{QSL}$ represents a slower dephasing speed limit, while a
shorter $\tau _{QSL}$ means a faster dephasing speed limit.

In Fig.~\ref{4.eps}, we plot the QSL time as a function of the
$s$-wave scattering length $a_{B}$ in BEC reservoir for different
regimes of the state distances between initial and target quantum
states, and different spatial dimensions of the BEC. In order to
consistent with the condition of dilute and weakly-interacting
gases, the scattering length can be tuned up to a maximum value
given by $a_{B,max}\approx3a_{Rb}$ for 3D BEC gases,
$a_{B,max}\approx2a_{Rb}$ for 2D case, and $a_{B,max}\approx a_{Rb}$
for 1D case by using Feshbach resonances, respectively
\cite{Haikka2011}. Here the initial state of the impurity atom is a
maximally coherent state with $x^2+y^2=1$ and $z=0$.  In
Fig.~\ref{4.eps} the color degree of freedom denotes the distances
between initial and target quantum states, and the dashed lines are
equal-value lines with the same distances.

Fig.~\ref{4.eps} shows that the QSL time can be controlled by changing the scattering length  and spatial dimensions of the BEC. From Fig. 4 we can see the following results. (1)  For  a given state distance $\mathcal{D}$  denoted by a dashed line, a larger
$a_{B}$ leads to a longer QSL time. Hence, the increase of nonlinearity
in BEC reservoir can prolong the DSL time of the impurity atom. And this
result holds for 1D,  2D, and 3D BEC reservoirs. (2) The larger  the state distance $\mathcal{D}$,  the longer the QSL time is. This implies that the DSL time  will be prolonged with the increase of the state distance. But it should be noted that,
for a given state distance, the DSL time could tend to be infinite by
increasing the scattering length $a_{B}$. This phenomenon can be
explained as a result of the stationary coherence of the impurity atom
induced by the nonlinear interaction of BEC reservoir, as shown in
Fig.~\ref{2.eps}. (3) Comparing the three subfigures in Fig.~\ref{4.eps} we can find that the QSL
time could also be extended by decreasing the effective dimension of
the BEC reservoir in the small-value regime of the state distance. Therefore, we can conclude that  the
enhanced nonlinear interaction and the lower dimension of BEC reservoir can
induce a longer DSL time. As the $a_{B}$ can be tuned via Feshbach
resonances, this would provide a practical way to prolong the
DSL time of the impurity atom.

%%%%%%%%%%%%%%%%%%%%%%%%%%%%%%%%%%%%%%%%%%%%%%%%%
\begin{figure}[tbp]
\centering
\subfigure{\includegraphics[clip=true,width=0.4\textwidth]{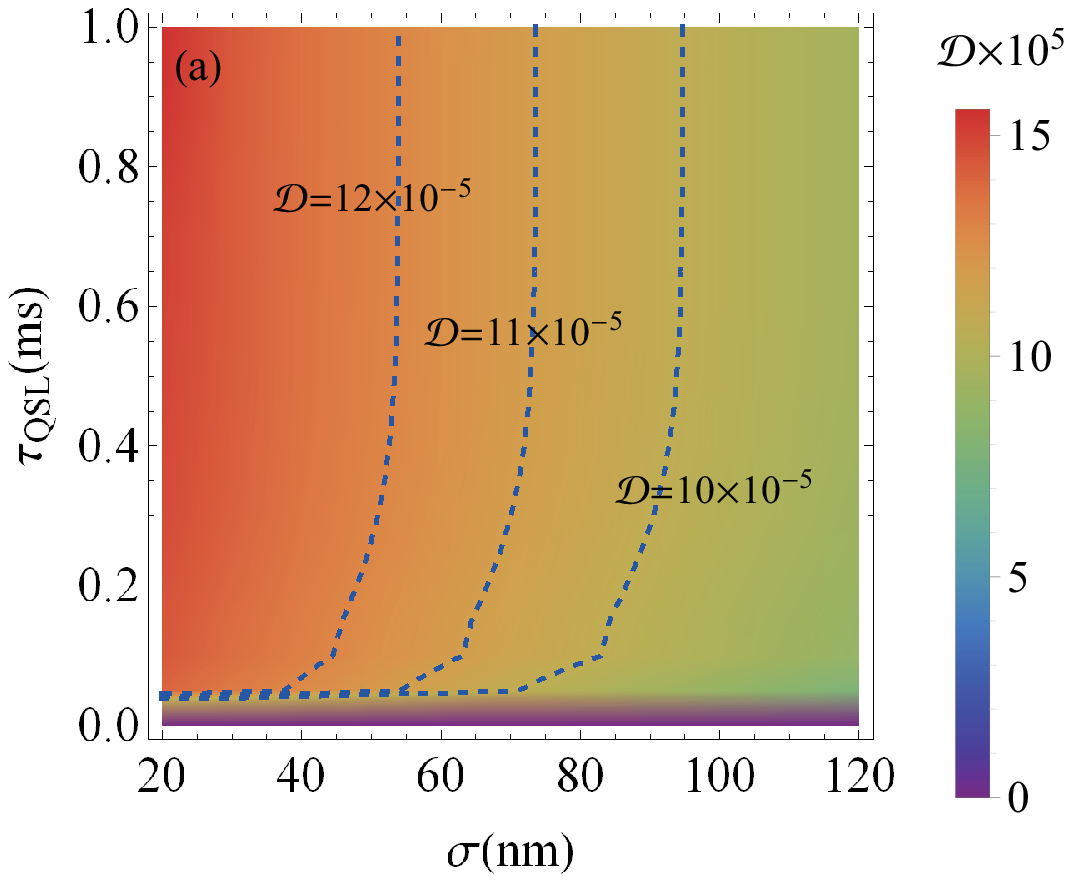}}
\subfigure{\includegraphics[clip=true,width=0.4\textwidth]{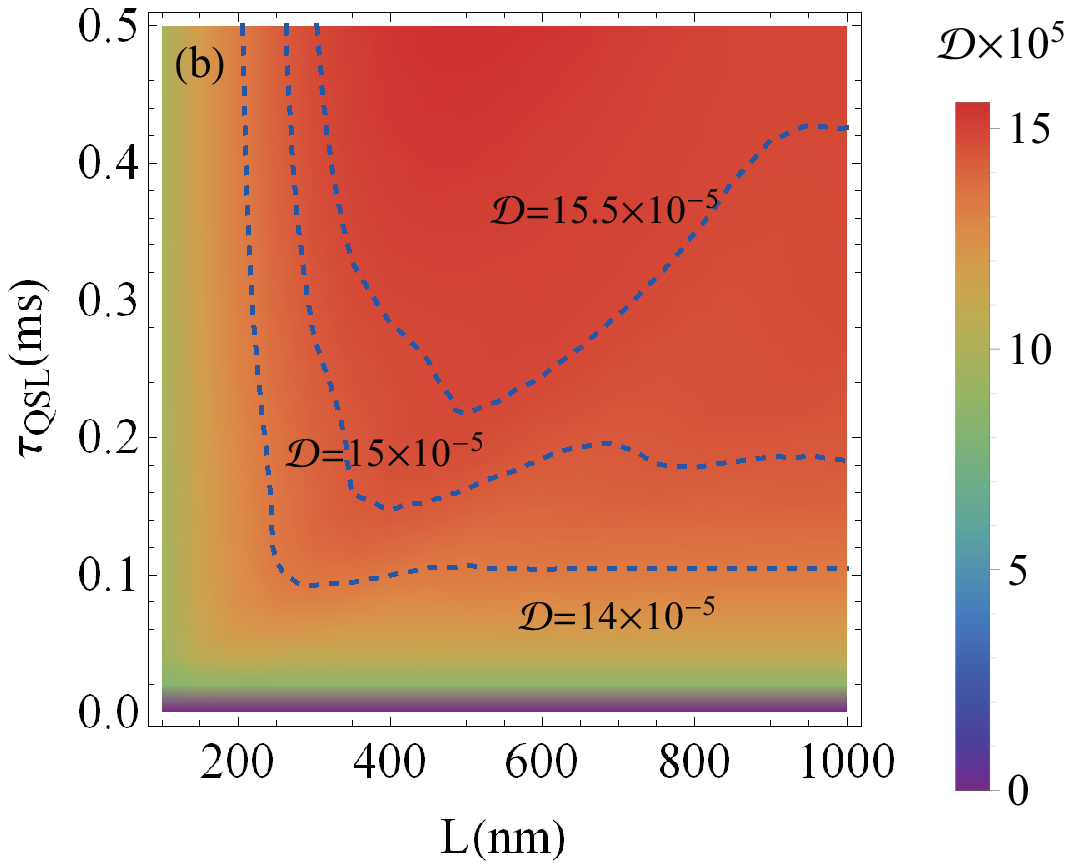}}
\caption{(Color online) The QSL time $\tau_{QSL}$ in 3D BEC
reservoir as a function of the trap parameter of double-well
$\sigma$ (the top subfigure) and the distance of two wells of
double-well $L$ (the bottom subfigure). The blue dashed contour
lines mark different given distances $\mathcal{D}$: from top to
bottom, the $\mathcal{D}$ takes the value of (a)
$\mathcal{D}\times10^{5}=12,11,10$ with $L=150nm$; (b)
$\mathcal{D}\times10^{5}=15.5,15,14$ with $\sigma=45nm$. The
initial-state of the impurity atom is prepared in a maximally
coherent state with $x^2+y^2=1$, $z=0$. And the s-wave scattering
length takes the value of $a_{B}=a_{Rb}=5.3nm$. Other parameters are
the same as in Fig.~\ref{2.eps}.}\label{5.eps}
\end{figure}
%%%%%%%%%%%%%%%%%%%%%%%%%%%%%%%%%%%%%%%%%%%%%%%%%%

Furthermore, Fig.~\ref{5.eps} presents the influence of two  spatial
parameters of the double-well potential $V_{A}(\textbf{r})$ ($\sigma$ and $L$)
on the DSL of the impurity atom for different regimes of the state distances in the case of the 3D BEC reservoir. Inspecting the
contour lines in Fig.~\ref{5.eps}, we can find that, for a given state
distance $\mathcal{D}$, the QSL time of the impurity atom $\tau_{QSL}$ in the 3D BEC
reservoir increases monotonously with increasing $\sigma$, while
decreases first and then oscillates with the increase of $L$. And
the manipulating mechanism is more efficient in the larger $\sigma$
regime or in the small $L$ regime. That is to say, the larger $\sigma$
and smaller $L$ induce a slower DSL. These results
can also be explained as a result of the effect of $\sigma$ or $L$
on the dephasing function $\Gamma(t)$, as shown in
Figs.~\ref{3.eps}. And we find similar results are also applicative
for the low-dimension cases. Anyway, above results provide us
another way to protect the coherence of the impurity atom in its BEC
reservoir.

The DSL manipulation of the impurity atom in BEC reservoir is
realized by controlling the spectral density of the BEC reservoir.
In order to explore  a more intuitive physical insight into the
effect of relevant  parameters, including $a_{B}$, $D$, $\sigma$ and
$L$, on the quantum dephasing process of the impurity atom, we
analyze the spectral density $J(\omega)$ in details in appendix A
\cite{Haikka2011,Hangleiter2015,Carole2014}. For qualitative
analysis, here we mainly consider the spectral density in the
low-frequency region, which plays a leading role in the long-time
decoherence process. In the non-interacting BEC reservoir
($a_{B}=0$), low-energy excitations with $\omega\ll\omega_{L}$ have
particle-like spectrum scaled as
\begin{eqnarray}
\label{3eq-9} J_{D}\left( \omega \right)  &\approx
&A_{D}\omega ^{D/2}e^{-\omega /\omega _{c}},
\end{eqnarray}
where the soft cut-off
frequency $\omega_{c}=\frac{1}{m_{B}\sigma^{2}}$  and three prefactors $A_{D}$ are given by Eq. (A8).

However, in the interacting BEC reservoir with $a_{B}\neq0$, the
low-energy excitations with $\omega\ll n_{D}g_{D}$ have phonon-like
spectrum scaled as
\begin{eqnarray}
\label{3eq-10} J_{D}\left( \omega \right)  &\approx
&B_{D}\omega^{D+2}e^{-\omega ^{2}/\tilde{\omega}_{c,D}^{2}},
\end{eqnarray}
where $\tilde{\omega}_{c,D} =\frac{\sqrt{2}c_{D}}{\sigma}$  with the speed of sound
$c_{D}=\sqrt{\frac{n_{D}g_{D}}{m_{B}}}$, and  three prefactors $B_{D}$ are given by Eq. (A10).

That is to say, for the free cases, the spectrum are sub-Ohmic,
Ohmic, and super-Ohmic in $1$D, $2$D and $3$D reservoirs,
respectively. However, the low-frequency spectrum  for the
interacting cases are always super-Ohmic, and the super-Ohmicity of
the spectral density is enhanced with the increase of dimensions of
the BEC reservoir. Thus, the slow down of the quantum dephasing
speed induced by increasing $a_{B}$ or decreasing $D$ can be
explained as a result of the change of the spectral density's
Ohmicity. And another interesting phenomenon is that the cut-off
frequency $\omega_{c}$ or $\tilde{\omega}_{c,D}$ can both be reduced
by increasing $\sigma$, which would also suppress the dephasing of
impurity induced by the high-frequency modes in the BEC reservoir.
Meanwhile, the dependence of the spectrum on the distance $L$ is
established through the relationship between the prefactors ($A_{D}$
or $B_{D}$) and $L$. From Eqs.~(A8) and~(A10), we can see that, in
the small region of $L$, the effective coupling constant is
proportion to $L^{2}$, which is the reason for the speedup of the
quantum dephasing of the impurity with the increase of $L$.

%%%%%%%%%%%%%%%%%%%%%%%%%%%%%%%%%%%%%%%%%%%%%%%%%%%%%%%%%%%%%%%%
\section{\label{Sec:5} Conclusions}
%%%%%%%%%%%%%%%%%%%%%%%%%%%%%%%%%%%%%%%%%%%%%%%%%%%%%%%%%%%%%%%%

In conclusion, we have studied  the DSL of a single impurity
atom immersed in a BEC reservoir with the impurity atom being in a double-well potential. We have obtained the DSL based on the quantum Fisher information formalism.
We demonstrated that the DSL of the impurity atom can be manipulated by engineering  the BEC
reservoir and  the impurity potential  within experimentally
realistic limits. It has been shown that the DSL can be controlled by
changing key parameters such as the scattering length,
the effective dimension of the BEC reservoir, and the spatial
configuration of the double-well potential.
In order to explore the physical mechanisms of controlling the DSL, we  have
analyzed the spectral density in details.  It has been revealed that the physical mechanisms of controlling the DSL at root of engineering
the spectral density of the BEC reservoir.

It is believed that these results of the present study may provide a
direct path towards engineering quantum dephasing speed of qubits in
nonlinear reservoirs, and would help to address the robustness of
quantum simulators and computers in a phasing damping channel
against decoherence \cite{Mukherjee2013,Iman2015,Carole2015}, which
may have implications in quantum cooling and quantum thermodynamics.
%%%%%%%%%%%%%%%%%%%%%%%%%%%%%%%%%%%%%%%%%%%%%%%%%%%%%%%%%%%%%%%%
\begin{acknowledgments}
This work was supported by the
National Natural Science Foundation of China under Grants Nos 11775075 and  11434011,  the Science
Foundation of Hengyang Normal University  under Grant No. 17D19, and the Open Foundation of Hunan
Normal University under Grant  No. QSQC1804.

\end{acknowledgments}

\appendix
\section{Derivation of the spectral density }

In this Appendix we derive the spectral density of  the impurity-BEC system.  The spectral density of the impurity-BEC system is formally given by
\begin{equation}
J\left(
\omega \right) \equiv \sum_{\mathbf{k}}\left\vert
\xi_{\mathbf{k}}\right\vert ^{2}\delta \left( \omega
-\omega_{\mathbf{k}}\right).
\end{equation}
In the continuum limit $\frac{1}{V}\sum_{\mathbf{k}} \rightarrow
\frac{1}{\left( 2\pi \right) ^{D}}\int d\Omega _{D}\int dk k^{D-1}$
with $\Omega _{D}$ being the surface of the unit sphere in $D$
dimensions, the spectral density can be rewritten by inserting
Eq.~(\ref{eq-3}) into the above equation,
\begin{eqnarray}
\label{eq-a1}
J\left( \omega \right)  &=&\frac{\eta _{D}^{2}n_{D}}{V}\sum_{\mathbf{k}}%
\frac{\epsilon _{\mathbf{k}}}{\omega _{\mathbf{k}}}e^{-k^{2}\sigma
^{2}/2}\sin ^{2}\left( \mathbf{k}\cdot \mathbf{L}\right) \delta
\left(
\omega -\omega _{\mathbf{k}}\right)  \nonumber\\
&=&\frac{\eta _{D}^{2}n_{D}}{\left( 2\pi \right) ^{D}}\int dkk^{D-1}f_{D}(kL)%
\frac{\epsilon _{\mathbf{k}}}{\omega _{\mathbf{k}}}e^{-k^{2}\sigma
^{2}/2}\delta \left( \omega -\omega _{\mathbf{k}}\right),\nonumber\\
\end{eqnarray}
where the angular integral is defined by
\begin{equation}
f_{D}(k L)=\int d\Omega _{D}\sin ^{2}\left( \mathbf{k}\cdot
\mathbf{L}\right),
\end{equation}
which can be easily calculated with the following results
\begin{eqnarray}
f_{1}(k L)&=& \sin ^{2}(kL), \hspace{0.5cm} f_{2}(kL)=\pi[1-J_{0}(2kL)], \nonumber\\
f_{3}(k L)&=&2\pi\left[1-\frac{\sin(2kL)}{2kL}\right],
\end{eqnarray}
where  $J_{0}(x)$ is the first kind of Bessel function .

Let $k(\omega)$ is the root of the equation $\omega
=\omega_{\mathbf{k}}$ in Eq.~(\ref{eq-2}),
\begin{eqnarray}
\label{eq-a2} k\left( \omega \right) &=& \sqrt{2m_{B}\left[
\sqrt{n_{D}^{2}g_{D}^{2}+\omega ^{2}}-n_{D}g_{D}\right] },
\end{eqnarray}
the spectral density in Eq. (A2) becomes
\begin{eqnarray}
\label{eq-a3} J\left( \omega \right)&=&\frac{\eta
_{D}^{2}n_{D}}{\left( 2\pi \right) ^{D}}\int d\omega _{\mathbf{
k}}k^{D-1}f_{D}(kL)\frac{\epsilon _{\mathbf{k}}}{\omega
_{\mathbf{k}}} \nonumber\\
&&\times e^{-k^{2}\sigma ^{2}/2}\left( \frac{d\omega
_{\mathbf{k}}}{dk}\right)
^{-1}\delta \left( \omega -\omega _{\mathbf{k}}\right)  \nonumber\\
&=&\frac{\eta _{D}^{2}n_{D}m_{B}}{\left( 2\pi \right) ^{D}}k\left(
\omega
\right) ^{D-2}f_{D}\left[ k\left( \omega \right) L\right] \frac{\epsilon _{%
\mathbf{k}}\left( \omega \right) e^{-k\left( \omega \right)
^{2}\sigma ^{2}/2}}{n_{D}g_{D}+\epsilon _{\mathbf{k}}\left( \omega
\right) },\nonumber\\
\end{eqnarray}
where the first partial derivatives of $\omega_{k}$ versus wave
vector $k$ can be derived from Eq.~(\ref{eq-2}), $\frac{d\omega
_{\mathbf{k}}}{dk} =\frac{k\left( n_{D}g_{D}+\epsilon
_{\mathbf{k}}\right) }{m_{B}\omega _{\mathbf{k}}}$.

As one can see from Eq.~(\ref{eq-2}), the dispersion relation is
much sensitive to the inter-atomic interaction strength $g_{D}$. In
the following we consider two extreme cases. In the free boson
reservoir with $g_{D}=0 (a_{B}=0)$, the quasi-particle energy tends
to $\omega =\frac{k^{2}}{2m_{B}}$.  In the low-frequency region with
$\omega\ll\omega_{L}$, the spectral density can be
reduced to the simple form
\begin{eqnarray}
\label{eq-a5} J_{D}\left( \omega \right)  &\approx
&A_{D}\omega ^{D/2}e^{-\omega /\omega _{c}},
\end{eqnarray}
where $\omega _{c} =\frac{1}{m_{B}\sigma ^{2}}$ and we have introduced three prefactors
\begin{eqnarray}
A_{1}&=&\frac{\eta_{1}^{2}n_{1}m_{B}^{3/2}L^2}{\sqrt{2}\pi}, \hspace{0.5cm} A_{2}=\frac{\eta_{2}^{2}n_{2}m_{B}^2L^2}{2\pi}, \nonumber\\
A_{3}&=&\frac{\sqrt{2}\eta_{3}^{2}n_{3}m_{B}^{5/2}}{3\pi^{2}},
\end{eqnarray}
where $n_D$ and $\eta_D$ are the number density of the D-dimensional
BEC and the interaction strength between impurity atoms and
D-dimensional BEC with the expressions given by Eqs.~(5) and~(6),
respectively. Making use of Eqs.~(5) and~(6), from Eq.~(A8) we can
see that the prefactors $A_D$ are dependent of the distance
parameter between two wells $L$ and the $s$-wave scattering length
$a_{AB}$ for BEC-impurity collisions.

Hence, in the non-interacting case, low-energy excitations of the
BEC reservoir have particle-like spectrum  given by Eq.~(A7) with
the soft cut-off frequency $\omega_{c}$
\cite{Haikka2011,Hangleiter2015}.  From Eq.~(A7) we can see that the
Ohmicity of the BEC reservoir is determined by the dimensions of the
BEC. The spectrum is sub-Ohmic, Ohmic, and super-Ohmic for 1D, 2D,
and 3D BEC reservoir, respectively.

Whereas for the interacting BEC reservoir with $g_{D}\neq 0$ $
(a_{B}\neq 0)$, in the low-frequency region with $\omega\ll
n_{D}g_{D}$, the dispersion relation changes to the phonon-like form
$\omega \approx c_{D}k$ with the speed of sound
$c_{D}=\sqrt{\frac{n_{D}g_{D}}{m_{B}}}$. Thus, the spectrum can be
approximately equal to
\begin{eqnarray}
\label{eq-a7} J_{D}\left( \omega \right)  &\approx
&B_{D}\omega^{D+2}e^{-\omega ^{2}/\tilde{\omega}_{c,D}^{2}}.
\end{eqnarray}
where $\tilde{\omega}_{c,D} =\frac{\sqrt{2}c_{D}}{\sigma}$ and we have introduced three prefactors
\begin{eqnarray}
B_{1}&=&\frac{\eta_{1}^{2}L^2}{4\pi g_{1}^{5/2}}\left(\frac{m_B}{n_1}\right)^{3/2},  \hspace{0.5cm}
B_{2}=\frac{\eta_{2}^{2}L^2}{8\pi g_{2}^3}\left(\frac{m_B}{n_2}\right)^{2},  \nonumber\\
B_{3}&=&\frac{\eta_{3}^{2}L^2}{12g_{3}^{7/2}\pi^{2}}\left(\frac{m_B}{n_3}\right)^{5/2}.
\end{eqnarray}
where $g_D$ is the interaction strength between inter-atomic
interaction in D-dimensional BEC with the expressions given by
Eq.~(4). Making use of Eqs.~(4)-(6), from Eq.~(A10) we can see that
the prefactors $B_D$  depends on not only the distance parameter
between two wells $L$ and the $s$-wave scattering length for
impurity-BEC collisions $a_{AB}$, but also the scattering length for
inter-atomic collisions $a_{B}$. Therefore, in the case of the
interacting BEC reservoir, low-energy excitations of the reservoir
have phonon-like spectrum given by Eq.~(A9). The spectral Ohmicity
of the BEC reservoir is always super-Ohmic for 1D, 2D, and 3D BEC
reservoir.

%%%%%%%%%%%%%%%%%%%%%%%%%%%%%%%%%%%%%%%%%%%%%%%%%%%%%%%%%%%%%%%%%%%%%%%%%%%%%%%%%%%%%%%

\end{document}